\documentstyle[astron,11pt,newpasp,twoside]{article}
\def\edcomment#1{\iffalse\marginpar{\raggedright\sl#1\/}\else\relax\fi}
\reversemarginpar

\begin{document}
\title{Particle Acceleration and Diffusion in Fossil Radio Plasma}
\author{Torsten A. En{\ss}lin}
\affil{Max-Planck-Institut f\"{u}r Astrophysik, Garching, Germany}

\begin{abstract}
The strong activity of radio galaxies should have led to a nearly
ubiquitous presence of fossil radio plasma in the denser regions of
the inter-galactic medium as clusters, groups and filaments of
galaxies. This fossil radio plasma can contain large quantities of
relativistic particles (electrons and possibly protons) by magnetic
confinement.  These particles might be released and/or re-energized
under environmental influences as turbulence and shock waves. Possible
connections of such processes to the formation of the observed sources
of diffuse radio emission in clusters of galaxies (the cluster radio
halos and the cluster radio relics) are discussed.
\end{abstract}

\section{The Cosmological Energy Budget of Radio Plasma}
The remnant of the radio lobes of a former active radio galaxy get
rapidly invisible for radio telescopes, justifying the name {\it
'radio ghosts'} for such patches of fossil radio plasma
(\cite{1999dtrp.conf..275E}). Radio ghosts should be an important
ingredient of the inter galactic medium (IGM). The present space
density of one or a few radio galaxies per galaxy cluster was exceeded
by orders of magnitude during the epoch of violent quasar activity
around $z = 2$. There are several routes to estimate the energy
released by earlier radio galaxies. One is to integrate the evolution
of the radio-luminosity-function, translated to jet-power via an
empirical jet-power--radio-luminosity relation, based on
equipartition energy densities of radio lobes. The result is
(\cite{2000A&A...360..417E})
\begin{equation}
E_{\rm jet} = 3\cdot 10^{66} \,{\rm erg\,Gpc^{-3}} \,(f_{\rm eq}/3)\,,
\end{equation}
where $f_{\rm eq}$ gives the ratio of true to equipartition energy
density in the radio plasma, which is likely bigger than one. This
amount of energy corresponds to $\sim 0.1$ keV per cosmic baryon, and
is therefore not negligible, since large-scale structure formation has
produced $\sim 1$ keV/baryon by gravitational infall.

An independent route is to use the Magorrian relation ($\eta_{\rm bh}
= M_{\rm bh}/M_{\rm bulge} = 0.002$) between the galaxy bulge ($M_{\rm
bulge}$) and central black hole ($M_{\rm bh}$) mass. Assuming an
efficiency of $\varepsilon_{\rm jet} = 0.1$ of rest mass to jet
energy conversion of the accretion process feeding the black hole
growth, one derives from the observed present day galaxy population
(\cite{1998AA...333L..47E})
\begin{equation}
E_{\rm jet} = 3\cdot 10^{67} \,{\rm erg\,Gpc^{-3}} \,(\eta_{\rm
bh}/0.002)\,(\varepsilon_{\rm jet} /0.1)\,.
\end{equation}
If this estimate is correct, this amount of energy could easily
provide the cosmic entropy-floor of 1 keV/baryon, which seems to be
required in order to explain the X-ray luminosity--temperature
relation of groups and clusters of galaxies.

Finally, a third way to estimate the energy in radio plasma is to
assume that the efficiencies of accretion discs in active galactic
nuclei (AGN) to produce X-rays and radio jets are the same. Assuming
further that the observed X-ray background results only from
unresolved AGN, one gets (\cite{1998AA...333L..47E})
\begin{equation}
E_{\rm jet} = 3\cdot 10^{67} \,{\rm erg\,Gpc^{-3}} \,(\varepsilon_{\rm
jet} /\varepsilon_{\rm X-ray})\,.
\end{equation}

This should illustrate the importance of fossil radio plasma for IGM
properties. A discussion of the possible influences of radio ghosts on
various aspects of extra-galactic astrophysics can be found in
\cite{1999dtrp.conf..275E} and \cite{Pune99}. Here, recent progress on
two of these aspects should be briefly reported: an attempt to
estimate the escape rate of relativistic particles from radio ghosts
(\cite{crdiff2000}), and a new model for the revival of of fossil
radio plasma (\cite{ensslin2000b}).

\section{Adiabatic Revival of Fossil Radio Plasma}

There is growing evidence that the so called {\it 'cluster radio
relics'} are tracers of shock waves in clusters of galaxies, as
proposed in \cite{1998AA...332..395E}. But these relics are rare
compared to the frequency of cluster merger shock waves. This and the
morphological connection between the relic 1253+275 in the Coma
cluster and the tails of the radio galaxy NGC 4789 indicate that the
locations which become radio luminous host fossil radio plasma. If
this is indeed the case, the mechanism energizing the electron
population is likely not Fermi-I shock acceleration. The reason for
this is that an environmental shock wave is only capable of an
adiabatic compression of the relativistic plasma. But due to the soft
equation of state, even adiabatic compression can lead to a
substantial energy gain of the electron population
(\cite{ensslin2000b}). If the fossil radio plasma was not too old, the
upper cutoff of the electron energy spectrum might be adiabatically
shifted to radio observable energies. Radio plasma younger than $\sim
0.1$ Gyr in the centers of galaxy clusters, and $\sim 1$ Gyr at
peripheral regions can be revived to radio emission in typical shock
waves. But this requires that the relativistic electrons are still
confined to the radio ghost.

\section{Particle Escape from Fossil Radio Plasma}

All the plasma of a radio lobe was injected into the IGM from a very
small region (the AGN) and expelled the IGM gas from a large
volume. No magnetic flux can therefore leak from the radio plasma into
the thermal environment, unless magnetic reconnection between these
two phases took place. If such reconnection happened, the
field topology could be partly opened, and relativistic particles
would be able to leave the blob of radio plasma along an inter-phase
magnetic flux tube. But also anomalous cross field diffusion might
lead to particle losses. Typical escape frequencies of 10 GeV 
particles from fossil radio plasma are estimated in
\cite{crdiff2000}. With the assumptions and parameters specified there
(field strength $B = 10\, \mu$G, characteristic scales of the
turbulence $l_B = \lambda_\| = l_\perp = \lambda_\perp = 10\,$kpc, and
a radio lobe diameter of 100 kpc) one gets a flux tube escape rate of
$\nu_{\|} \approx 2.0\,{\rm Gyr}^{-1}\, \eta_s/\varepsilon_0$, and a
perpendicular escape rate of $\nu_{\perp} \approx 0.35\,{\rm
Gyr}^{-1}\, \delta_0/\varepsilon_0$. $\delta_0\le1$ gives the
ratio of turbulent to total magnetic energy on the turbulence
injection scale $l_{\rm B}$, $\varepsilon_0 \le 1$ describes the
efficiency of particle pitch angle scattering, and $ \eta_s\ll 1$ is
the fraction of the lobe's surface threatened by inter-phase magnetic
flux tubes.  Particle losses are therefore slow even on cosmological
time-scales.

These estimates assume a Kolmogorov turbulence cascade from the large
scales, which are important for the anomalous particle transport
across the field lines, down to the microscopic scales, responsible for
pitch angle scattering and therefore fixing the parallel diffusion
coefficient. If the shape of the spectrum is different, the results
change. E.g. the presence of much stronger large scale turbulence can
lead to the regime of enhanced anomalous diffusion. 

If a cluster merger event suddenly injects large-scale turbulence the
radio plasma can get temporary transparent even for low energy
particles.  Let's assume that the merger produces turbulent flows with
velocities of $v_T \approx 1000\,$km/s on a scale of $l_T \approx
100\,$kpc which increase the turbulent magnetic energy density on
large scales by a factor of $X_T = 30$ from initially $\delta_0 =
0.01$. For roughly an eddy turnover time $\tau_T = l_T / v_T \approx
100\,$ Myr the small scale turbulence is not increased. During this
period an enhanced anomalous cross field escape should allow roughly
$\tau_T \,\nu_\perp \,X_T^2 \approx 30 \%/ \varepsilon_0$ of the 10
GeV particles initially confined in the radio ghost to escape. The
losses would even be much higher, if the pitch angle scattering
efficiency is low ($\varepsilon_0 \ll 1$). The total loss of particles
escaping along inter phase flux tubes is $\tau_T \,\nu_\| \,X_T
\approx 6 \%\, (\eta_{\rm s}/0.01) / \varepsilon_0 \ll 1$, and likely
negligible if $\eta_{\rm s}$ is very small.

The release of relativistic protons from fossil radio plasma into the
gaseous IGM might help to explain the observed {\it `cluster radio
halos'} via hadronic secondary electron production
(\cite{1999dtrp.conf..275E,Pune99}).

\acknowledgments I thank Luigina Feretti for the invitation to the
Joint Discussion Session at the IAU General Assembly 2000. I
apologize for ignoring here many relevant articles from other
authors due to the lack of space.


\end{document}